# Public Opinion and The Rise of Digital Minds: Perceived Risk, Trust, and Regulation Support

Justin B. Bullock[1,2,3,4]
Janet V.T. Pauketat[5][0000-0003-3280-3345]
Hsini Huang[6,7][0000-0001-7546-1096]
Yi-Fan Wang[8][0000-0003-1409-9753]
Jacy Reese Anthis[5,9,10][0000-0002-4684-348X]

[1]Americans for Responsible Innovation
jbull14@gmail.com
[2]Convergence Analysis, Berkeley, CA, USA
[3]University of Washington, Seattle, USA
[4]Texas A&M University, College Station, TX, USA
[5]Sentience Institute, New York, NY USA
[6]Leiden University, The Hague, Netherlands
[7]National Taiwan University, Taipei, Taiwan
[8]National Cheng Kung University, Tainan, Taiwan
[9]University of Chicago, Chicago, IL, USA
[10]Stanford University, Stanford, CA, USA

**Abstract.** Governance institutions must respond to societal risks, including those posed by generative AI. This study empirically examines how public trust in institutions and AI technologies, along with perceived risks, shape preferences for AI regulation. Using the nationally representative 2023 Artificial Intelligence, Morality, and Sentience (AIMS) survey, we assess trust in government, AI companies, and AI technologies, as well as public support for regulatory measures such as slowing AI development or outright bans on advanced AI. Our findings reveal broad public support for AI regulation, with risk perception playing a significant role in shaping policy preferences. Individuals with higher trust in government favor regulation, while those with greater trust in AI companies and AI technologies are less inclined to support restrictions. Trust in government and perceived risks significantly predict preferences for both soft (e.g., slowing development) and strong (e.g., banning AI systems) regulatory interventions. These results highlight the importance of public opinion in AI governance. As AI capabilities advance, effective regulation will require balancing public concerns about risks with trust in institutions. This study provides a foundational empirical baseline for policymakers navigating AI governance and underscores the need for further research into public trust, risk perception, and regulatory strategies in the evolving AI landscape.

**Please cite as** Bullock, J.B., Pauketat, J.V.T., Huang, H., Wang, Y-F. & Anthis, J.R. (2025). Public opinion and the rise of digital minds: Perceived risk, trust, and regulation support. *Public Performance and Management Review.* https://doi.org/10.1080/15309576.2025.2495094

**Note.** This is the author's preprint version of the manuscript and may not exactly match the version of record.

# 1 Introduction

Good governance requires risk management and trust. One increasingly important area for public management and public governance is the regulation of artificial intelligence (AI). In addition to the concerns of corporate and governmental stakeholders, good AI governance requires understanding the public's perceptions of technological risks, and their trust of the institutional and technological elements of the complex AI ecosystem for generative AI. Generative AI development has been fast-paced since the advent of ChatGPT in 2022, with no indication of slowing down. This fast-paced technological change raises critical concerns about the risks of AI to civil society and beliefs about whether companies and governments can meaningfully manage generative AI. Despite this, empirical research has neglected studying the relationship between the public's perceptions of AI risks and their trust in AI governance, industry, and systems. Establishing a baseline to evaluate public opinion on AI risk, trust of the AI ecosystem, and support for AI governance strategies is necessary before fine tuning regulations to mitigate societal risks of fast-paced technological change. We establish this empirical baseline with nationally representative USA data from the 2023 Artificial Intelligence, Morality, and Sentience (AIMS) online survey supplement (Pauketat, Bullock, et al., 2023; Pauketat, Ladak, et al., 2023) to offer critical insight into public opinion on AI and AI governance.

AI regulation issues capture scholars' and governments' attention because of the substantial risks associated with this technology. Although AI can boost governmental performance in public service design and delivery, this technology threatens various societal aspects, such as privacy infringement, racial discrimination, social surveillance, and insufficient explainability and accountability (Yuan & Chen, 2025; Nieuwenhuizen et al., 2024; Zuiderwijk et al., 2021). Many countries enact regulations of personal data protection and AI-enabled systems to minimize the risks and harms of this technology (Barkane, 2022; Cath, 2018; De Hert & Bouchagiar, 2022; Laux et al, 2024; Tan & Taeihagh, 2021). However, studies that investigate how citizens perceive and evaluate these AI regulatory policies remain scarce, except for a few recent explorations on why citizens support AI regulations (Heinrich & Witko, 2024; O'Shaughnessy et al., 2023).

Our theoretical contribution lies in deepening our understanding of how public opinion on trust and AI risk could shape policy preferences for different strengths of AI regulatory approaches. Citizens might support AI regulations but have different preferences on the strength and density of regulatory policies (e.g., soft or strong regulations). Also, some significant AI research streams, such as trust and perceived risks from citizens' perspectives, demonstrate that citizens' perceptions of AI might affect their acceptance of AI applications in public governance (Mei & Zheng, 2024; Lin et al., 2021; Wang et al., 2023; Wang et al., 2024). Based on this logic, perceived trust and risks may affect citizens' support of AI regulations.

The need to evaluate and develop diverse regulatory strategies to cope with AI risks is evident to expert stakeholders, but public opinion on AI developments and policy tools that manage generative AI is unclear, especially as use of generative AI systems like ChatGPT appears increasingly popular. Given generative AI's unknown risks, different regulations along a continuum of severity from relatively soft (e.g., slowing down developments with non-mandatory means) to relatively strong (e.g., banning developments and applications) may be necessary, and may be supported by the public. The policy-making process inevitably would be reviewed by the societal and political actors, such as politicians, bureaucrats, citizens, and interest groups, thus this paper investigates the public's perception of advanced AI risks, their trust of various elements of the AI ecosystem, and their support for soft and strong AI governance. Specifically, we ask (1) to what extent do U.S. Americans perceive the development of advanced AI technologies to be risky, (2) to what extent are the government, corporate, and AI systems themselves trusted, and (3) what are the relationships between perceived risk, trust, and policy preferences for AI risk mitigation?

## 2 Literature Review

### 2.1 AI Regulations

Many governments worldwide have introduced AI laws, rules, and policies to govern relevant risks, harms, and boost potential benefits, but their approaches often differ. For instance, the policies of the US government focus mostly on accelerating the development of AI innovations and technologies (Chen & Gasco-Hernandez, 2024), while the European approach emphasizes the necessity of legislation (Cath, 2018) to alleviate potential AI-led harms, injustice, and discrimination (De Hert & Bouchagiar, 2022). In 2024, the European Parliament approved the AI Act that stresses the strict regulation of AI-enabled realtime biometric surveillance and scraping facial recognition from Internet or CCTV footage based on a risk-level framework (Laux et al 2024; Barkane, 2022). Bode and Huelss (2023) discuss how the EU interacts with technology industries to enact regulations for military AI systems. In Singapore, the government utilizes regulatory sandboxes to allow companies to experiment with AI-enabled services on a small scale while avoiding unintended consequences (Tan & Taeihagh, 2021). The many real cases of AI applications have raised concerns and attention to regulatory policies to manage AI challenges and facilitate advantages.

These governmental strategies impact the public as well as AI industry stakeholders, but public opinion on AI regulation, in general, is not well understood, despite it being a well-established antecedent to policy-making. The general public has opinions on AI regulation issues. For example, Heinrich and Witko (2024) found that framing effects of AI automation on job replacement increased support for AI regulation amongst people with more knowledge about technology. In Singapore, the public's acceptance of autonomous AI systems has been linked to perceived risk and trust in the government to regulate (Pande & Taeihagh, 2024). Further, good AI governance requires a holistic incorporation of citizens' attitudes toward AI regulation in addition to macro-comparisons of regional, governmental, and political strategies. By asking citizens if they support the regulation of AI, policy makers can have a more comprehensive understanding of the tradeoffs that citizens make between trusting AI to benefit them, and feeling threatened by AI developments when deciding whether or not to support government intervention (O'Shaughnessy et al., 2023). Public opinion on AI, especially the effects of perceived risk and trust on regulation preferences, is another crucial lens through which to analyze AI regulation.

### 2.2 Public Preferences for the Regulation of Generative AI

Generative AI systems like large language models (LLMs) or so-called "digital minds" use deep learning techniques trained on large amounts of data (Anthis, 2023; Fui-Hoon Nah et al., 2023). Users of LLM-based chatbots converse with this new type of GenAI to receive answers to their questions, generate computer code, draft templates for various purposes, and even simulate complex agent-based interactions (Fui-Hoon Nah et al., 2023; Salah et al., 2023). Such generative AI systems are considered useful for knowledge-creation or sense-making tasks for organizational leaders to improve knowledge management (Korzynski et al., 2023). However, the many technical attributes of the generative AI systems raise privacy, accountability, moral and psychological concerns, in particular concerns about AI's social biases, discrimination, and misinformation (Jung & Camarena, 2024; Ladak et al., 2023; Manoli et al., 2025; Salah et al., 2023). Those above-mentioned uncertainties and risks of AI could affect citizens and their social life, which requires more research attention. However, citizens' attitudes and policy preferences towards generative AI systems like ChatGPT are understudied.

There is a wide range of policy tools governments can use to manage AI's development pace, risks, and negative consequences. We divide these policy choices into stronger (i.e., immediate full prohibitions) and softer (i.e., intervention in technology development) regulatory policies. By stronger regulations, we mean immediate bans or broadly prohibiting AI use. Stronger AI regulation would directly stop AI companies from developing and innovating in the generative AI space. For instance, the EU AI Act bans real-time biometric surveillance for identification purposes, manipulative AI, and untargeted scraping of facial images from Internet and CCTV footages (Barkane, 2022). Some U.S. local governments also prohibit using facial recognition technology and predictive algorithms in policing (Fountain, 2022). On the other hand, softer regulations are defined as government programs aiming to limit some AI applications or slow down technology development without direct enforcement. At the time of this study, for instance, the Japanese government is leaning toward softer regulations toward AI, to avoid limiting the development of new innovations by regulating them too strongly, too early (Warren et al., 2024).

## 2.3 Institutional Trust

To understand citizens' attitudes toward government policymaking, two types of institutional trust are often discussed: trust in government and trust in industry. Institutional trust, in which social institutions are accepted by individuals and organizations as facts, can affect citizens' perceptions of social objects (Thomas, 1998). When citizens consider AI-relevant policies, they may rely on their trust in the institutional environment surrounding the development of generative AI, namely trust in the government (rule-making) and trust in the companies (developing) generative AI (Williamson, 1993). Separating these two arms of institutional trust is valuable given their separate roles in the AI governance ecosystem (Bullock et al., 2023). This section discusses the roles of government trust and industry trust in AI regulation.

### 2.3.1 Government Trust

Government trust affects citizens' attitudes toward AI policies. The government serves as a regulator, and citizens' trust in the regulator determines how the public sector implements regulatory policies. That is, if citizens' trust the regulator, then they are more likely to follow and implement the policies as intended. When citizens have more trust in the government, the public sector can enforce regulatory policies more successfully (Six & Verhoest, 2017; Schmidt et al., 2018). For instance, with increased trust in the government, citizens support environmental policies even when those policies pose burdens to people, such as high taxes and regulations (Harring & Jagers, 2013; Huber & Wicki, 2021; Kallbekken & Sælen, 2011; Kitt et al., 2021), and bans on energy applications receive more support from civil society (Kulin & Johansson Sevä, 2021). Also, individuals' trust in the government substantially affects their attitudes toward nuclear policy, such as nuclear plant building or prohibition (Nam-Speers et al., 2023). Similarly, trust in incumbent politicians facilitates their support of fossil fuel taxes (Fairbrother et al., 2019).

Like in the environmental protection and nuclear energy contexts, generative AI brings tremendous benefits and substantial risks to society. For example, although AI systems increase efficiency, effectiveness, and profits to various sectors, this technology raises concerns about privacy, discrimination, job replacement, and surveillance (Zuiderwijk et al., 2021). Further, the AI computing process requires a large amount of energy, exacerbating climate change and the greenhouse effect (Coeckelbergh, 2022). Policies to regulate generative AI development is similarly a controversial issue, so trust in the government could predict AI regulation policy preferences. We propose the first cluster of research hypotheses below:

*Hypothesis 1a: Government trust increases the support of softer AI regulation.*
*Hypothesis 1b: Government trust increases the support of stronger AI regulation.*

### 2.3.2 AI Industry Trust

In addition to trusting the government, trusting AI companies is another determinant of policy support. Trust in regulated industries shapes citizens' attitudes toward regulatory policy (Six & Verhoest, 2017). When individuals trust an industry, they prefer to positively evaluate information about companies (Cvetkovich et al., 2002). Also, trust in the market decreases individuals' support of state-oriented welfare policy (Edlund & Lindh, 2013), meaning that citizens who believe market mechanisms can address social issues are less likely to support government intervention. Based on this logic, individuals confident in a particular industry tend to oppose regulations on that industry. For instance, trust in the fuel industry negatively affects citizens' support of gasoline supply regulations (Rhodes et al., 2017). However, when individuals distrust industry, they prefer the government to regulate that industry's commercial activities. For example, lower trust in industry increases support of fossil fuel consumption regulations (Dietz et al., 2007). Also, distrust in commercial actors increases support of punitive policies for environmental protection (Harring, 2018). Hence, trust in industry or private companies affects regulatory policy support. Thus our second cluster of research hypotheses are:

*Hypothesis 2a: AI Industry trust decreases the support of softer AI regulation.*
*Hypothesis 2b: AI Industry trust decreases the support of stronger AI regulation.*

### 2.3.3 Trust in AI Technology

Trust in AI technologies serves as a critical element of AI-enabled public service delivery. Given the high level of risks and uncertainties, trust determines whether individuals are willing to use AI or accept information and actions from this technology. For instance, Wang et al. (2023) suggest that citizens' intention to follow travel recommendations for COVID-19 is determined by trust in the AI-generated information. Similarly, the perceived transparency of AI determines citizens' trust in AI systems (Grimmelikhuijsen, 2023; Wang et al., 2023). Public employees who regard AI as positive and beneficial are more likely to support the adoption of this technology in public organizations (Ahn & Chen, 2022). Although, organizational position may also affect employees' trust and perceived value in AI. Empirically, public managers are found to be more likely to accept and support AI as the decisional support system than non-managerial staff in the public sector (Huang et al., 2022).

Additionally, citizens' trust in AI-enabled applications is likely to be influenced by policy domains in which AI is applied. For example, Gesk and Leyer (2022) found that citizens prefer AI in the context of general policy in which the majority of people are affected without having to directly request regulation, but they value humans more in specific policy contexts in which only a few people are affected and individuals have requested direct intervention. Second, policy context shapes trust in AI systems. Aoki (2020) suggests that Japanese residents trust AI more in waste collection than parental support contexts. Also, for tax-relevant policy, citizens trust human agents more than machine intelligence (Ingrams et al., 2022). Finally, Laux et al. (2023) reviewed the European Commission's AI Act and argue that the connections between regulations and AI trustworthiness remain unclear. Policy characteristics and trust in AI are intertwined and the effect of trust in AI on policy preferences has been understudied relative to the effect of policy context on trust in AI.

Notably, previous research highlights trust in AI as a crucial outcome in human-AI interaction. Scholars identify factors affecting citizens' trust in AI, such as transparency, explainability, policy context, privacy concerns, and fairness issues (Aoki, 2020; Chen et al., 2023; Grimmelikhuijsen, 2023; Lin et al., 2021; Wang et al., 2023). However, limited research discusses the impacts of trust in AI as a predictor of citizens' perceptions, intentions for AI use, and policy preferences. The few exceptions suggest that trust in AI strengthens individuals' willingness to follow the recommendations of an AI system (Lin et al., 2021; Wang et al., 2023). Moreover, the existing literature on digital tools and governance provides a lens to explore the impacts of trust in AI. For instance, citizens' trust in e-government improves their support of the government's investment in digital technologies (Horsburgh et al., 2011). Also, citizens using more e-government services are more likely to trust overall administrative processes in the government (Tolbert & Mossberger, 2006). When citizens trust public ICT applications, they are more likely to support using digital technologies, such as generative AI, in the public sector. Therefore, the third cluster of research hypotheses is:

*Hypothesis 3a: Trust in AI technologies decreases the support of softer AI regulation.*
*Hypothesis 3b: Trust in AI technologies decreases the support of stronger AI regulation.*

## 2.4 Perceived AI Risk

Governments around the world are keen to introduce AI machines and systems to improve public service delivery. European countries use AI to manage service delivery progress, boost policy implementation qualities, improve communications with citizens and commercial entities, and innovate new public policies (van Noordt & Misuraca, 2022). A report of the Dutch National Court of Audit released in October 2024 indicates that 58% of central government organizations had experience with AI (Netherland Court of Audit, 2024). Local governments within the U.K. have adopted virtual assistants and predictive analytics to streamline interactions with external actors and improve internal information and task management (Vogl et al., 2020).

The AI Use Case Inventory (https://ai.gov/ai-use-cases/) is a portal that records AI projects implemented by the Federal government in the United States. One objective of these AI applications is to increase the effectiveness and efficiency of public service delivery (Young et al., 2019). Expectations are to have automated AI systems that could assist public organizations in strengthening policy planning and implementation, AI-enabled chatbots to facilitate citizen communication to reduce processing time and provide customized responses (Androutsopoulou et al., 2019; Aoki, 2020). The current development of AI has been shown to benefit governments and citizens by improving the efficiency and effectiveness of public governance.

However, this technology also raises particular concerns about its harm to society and bureaucracy. AI could threaten ethical principles from racial discrimination, moral dilemmas, and the misalignment between human and machine judgments (Bullock, 2019; Wirtz et al., 2020; Young et al., 2019; Zuiderwijk et al., 2021). As Young et al. (2021) indicate, adopting AI increases the risk of administrative evil at individual, organizational, and institutional levels. AI harms include inscrutability, automation and quantification biases, the misconfiguration between technological and organizational values, the overreliance on AI, and the lack of AI performance evaluation and testing (Young et al., 2021).

Citizens' concerns over the many uncertainties around AI could come from insufficient transparency, social inequality, privacy issues, and job replacement effects. First, when data sources and algorithms in AI-based information processing remain ambiguous for users and citizens, individuals can rarely understand data collection and processing mechanisms in this technology, resulting in low perceived transparency from citizens (Chen et al., 2023; Grimmelikhuijsen, 2023; Grimmelikhuijsen & Meijer, 2022; Zuiderwijk et al., 2021). Second, AI raises citizens' fairness concerns because of the nature of surveillance. For instance, police departments utilize facial recognition technology to address human trafficking, target risky populations, and track potential criminal events. At the same time, these AI tools suffer from inaccurate identifications based on gender, age, and race (Fountain, 2022). This drawback results in inequitable treatment among different social groups, exacerbating existing social biases and inequality (Chen et al., 2023; Fountain, 2022; Moon, 2023). Third, AI applications threaten citizens' personal information and privacy protection. When the government uses AI to collect and analyze sensitive personal information to predict human actions and behaviors, citizens face significant privacy concerns from public AI applications (Grimmelikhuijsen & Meijer, 2022; Lin et al., 2021; Saura et al., 2022; Wang et al., 2023; Zuiderwijk et al., 2021). Finally, individuals are concerned about the replacement effects coming from an increase in the adoption of AI systems. Public organizations and commercial entities are increasingly adopting AI to automate various tasks, so public employees and citizens perceive a threat of job replacement (Wirtz et al., 2020). Public employees may regard AI as their competitors rather than as their assistants (Ahn & Chen, 2022). Therefore, individuals' have substantial concerns about the risks of AI systems to themselves and society.

One motivation for regulatory policy support is that individuals perceive risks, harms, and concerns about specific social problems. For instance, when citizens are concerned about environmental issues, they are more likely to support softer environmental protection policies, such as taxes and subsidies (Davidovic et al., 2020; Fairbrother, 2016). Also, climate change concerns increase citizens' preference for stronger regulations such as energy bans (Kulin & Johansson Sevä, 2021). Moreover, individuals with more concerns about nuclear energy are less likely to support adopting this technology (Stoutenborough et al., 2013) and perceived threats impede citizens from supporting the development of nuclear power (Ho & Chuah, 2021). Following this logic, individuals may be more likely to support regulations when they perceive AI as harmful. For public employees, their acceptance of generative AI in public administration depends on their evaluation of the safety and efficacy of this technology. If this technology benefits society, bureaucrats may prefer to adopt AI in public organizations (Ahn & Chen, 2022). However, the various AI risks and harms of generative AI, including privacy issues, unfairness, and unexpected consequences (Wang et al., 2023; Wirtz et al., 2022; Wirtz et al., 2020; Young et al., 2021) may trigger citizens' concerns about this technology that increase support for governmental regulation. Hence, the fourth cluster of research hypotheses is:

*Hypothesis 4a: Perceived AI risks increase the support of softer AI regulation.*
*Hypothesis 4b: Perceived AI risks increase the support of stronger AI regulations.*

Figure 1 integrates our four clusters of hypotheses into a theoretical model that we use to frame our foundational, exploratory investigation of the effects of trust and perceived risk on regulatory policy preferences in the context of generative AI in the USA.

**Figure 1. Theoretical Model.**

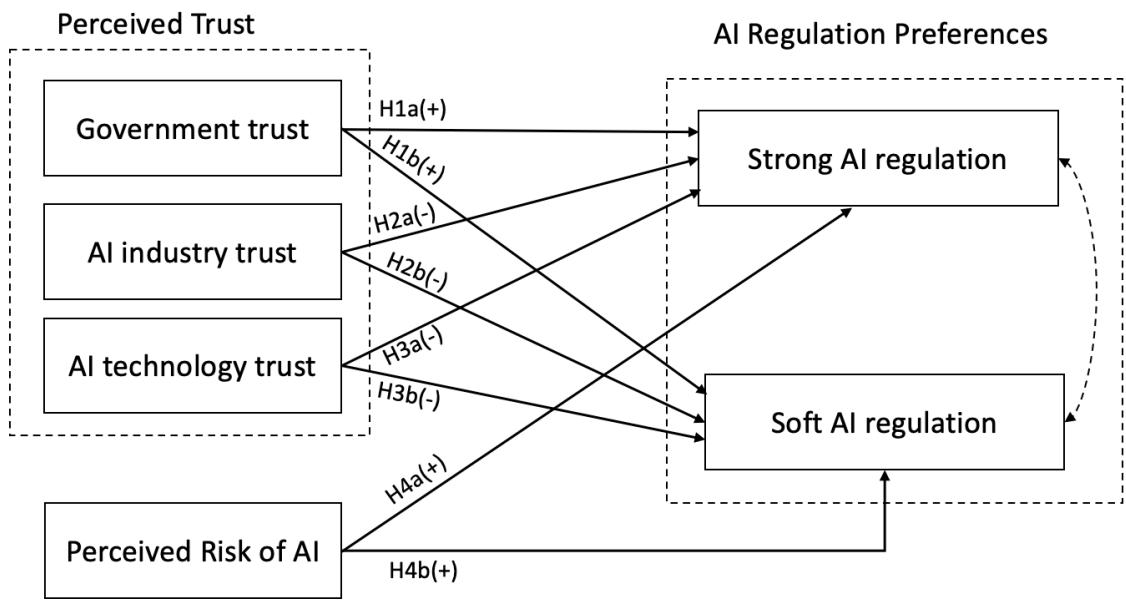

*Note.* This is the theoretical model for our empirical study of the predictive effects of trust and risk on AI regulation preferences.

## 3 Data and Methods

We employed nationally representative data from the 2023 Artificial Intelligence, Morality, and Sentience (AIMS) online survey supplement ($N = 1,099$) of U.S. adults' attitudes towards near-term developments in AI technologies (Pauketat, Bullock et al., 2023; Pauketat, Ladak et al., 2023) to test our hypotheses on an existing dataset. While we were not able to more rigorously test our hypotheses through a preregistered study, this provides a baseline assessment of the associations present in public opinion. The AIMS survey was conducted May–July 2023 with iSay/Ipsos, Dynata, Disqo, and other leading sample panels and the data is openly available on Mendeley Data. The sample was recruited according to the 2021 census estimates from the American Community Survey (ACS) for age, education, race/ethnicity, gender, income, and region. The sample demographics and ACS estimates are in Table 1.

The AIMS survey methodology was preregistered and is available on the Open Science Framework (OSF; https://osf.io/7p2wt/). In the AIMS survey, a variety of terms were used to describe AI technologies (e.g., "robots/AIs," "AI systems," "AI," "LLMs," "chatbots"). Many terms can, and have been, used in AI, HCI, and HRI research. We believe this variation in wording serves to enhance the robustness of the research given the diversity of technologies and ways of conceptualizing these systems. In this paper, we examined institutional trust in the government and AI companies to regulate and control AI, trust in AI technologies, and risk perceptions associated with AI as the predictors of support for policies that slow down AI development and ban certain advanced AI technologies. In our analyses, we controlled for the effects of gender, political orientation, race/ethnicity, income, education, and frequency of exposure to AI systems and news or stories about AI. Exposure frequency was the average of responses to two items measured on 0 (never) to 5 (daily) Likert-type scales.

**Table 1. AIMS Survey Supplement Demographics.**

| | ACS 2021 | AIMS 2023 | | ACS 2021 | AIMS 2023 | | ACS 2021 | AIMS 2023 |
|---|---|---|---|---|---|---|---|---|
| **Age** | | | **Ethnicity / Race** | | | **Education** | | |
| 18 – 34 | 28.3% | 27.8% | American Indian/Alaskan Native | 0.4% | 1.1% | Less than high school diploma | 10.4% | 10.3% |
| 35 – 54 | 33.2% | 33.3% | Asian/Hawaiian/Pacific Islander | 6.1% | 6.4% | High school diploma | 26.9% | 26.4% |
| 55 – 100 | 38.5% | 38.8% | Black, Non-Hispanic | 11.1% | 11.3% | Some college, no degree | 20.6% | 18.7% |
| | | | Hispanic | 17.1% | 17.3% | Associate's degree | 8.6% | 9.5% |
| **Gender** | | | Mixed Racial Background/Other | 4.0% | 3.8% | Bachelor's degree | 20.8% | 22.5% |
| Male | 48.8% | 48.3% | White, Non-Hispanic | 61.3% | 60.1% | Postgraduate degree | 12.7% | 12.7% |
| Female | 51.2% | 51.7% | | | | | | |
| | | | **HH Income** | | | | | |
| **Region** | | | $2,500 - $24,999 | 10.7% | 12.3% | | | |
| Northeast | 17.5% | 16.8% | $25,000 - $49,999 | 16.6% | 16.3% | | | |
| Midwest | 20.7% | 20.9% | $50,000 - $74,999 | 16.7% | 16.7% | | | |
| South | 38.1% | 38.4% | $75,000 - $99,999 | 14.0% | 14.0% | | | |
| West | 23.8% | 23.8% | $100,000+ | 42.0% | 40.8% | | | |
| *Note.* The ACS 2021 percentages are the most recent demographic estimates for the USA. | | | | | | | | |

## 3.1 Key Measures

Table 2 shows the descriptive statistics for the various trust measures, perceived AI risk, and policy preference items. Table 3 shows the descriptive statistics, correlations, and reliability and factor checks for the composite variables that we formed and analyzed in the statistical models. The last three columns of Table 3 indicate that most of the major composite variables have Cronbach's alpha scores greater than 0.6 and pass the Kaiser Criterion with eigenvalues greater than one using the principal component factor method, verifying content validity. We also employ the Harman's one-factor test as a simple check for the common method bias, which is usually a concern for cross-sectional surveys collecting the dependent and independent variables at the same time. By running a principal factor analysis with all the key variables (both DVs and IVs), we find that the total variance extracted by one factor accounts for 37%, which is lower than the 50% threshold, suggesting that common method bias is less of a concern. While these tests are not definitive in refuting common source bias as a concern, they provide suggestive evidence and are in line with the field's norms for addressing common source bias concerns in survey research.

**Table 2. Descriptive Statistics for the AIMS Survey Items.**

| Predictor Variable | Items | *M* (*SD*) | Outcome Variable | Items | *M* (*SD*) |
|---|---|---|---|---|---|
| Government Trust | Trust: Government | 3.09 (1.77) | Slowdown Support | Public Campaigns | 4.88 (1.84) |
| | Effective Regulation | 4.79 (1.82) | | Support Regulation | 4.90 (1.84) |
| | Regulation Power | 4.84 (2.47) | | Oppose Regulation | 3.40 (1.96) |
| Industry Trust | Trust:  Companies | 3.42 (1.72) | Ban Support | AGI | 4.66 (1.98) |
| | Creator Safety | 3.19 (2.40) | | Data Centers | 4.64 (1.94) |
| | Creator Control | 4.84 (2.47) | | Sentience | 4.82 (1.91) |
| AI Trust | Chatbots | 3.75 (1.84) | | | |
| | LLMs | 3.99 (1.78) | | | |
| | Robots | 3.74 (1.87) | | | |
| | Game-Playing AI | 4.15 (1.85) | | | |
| Perceived Risk | Extinction | 3.86 (2.03) | | | |
| | Personal Harm | 4.39 (1.93) | | | |
| | USA Harm | 4.94 (1.80) | | | |
| | Future Harm | 5.07 (1.75) | | | |

**Table 3. Descriptive Statistics and Correlations.**

|  | 1. | 2. | 3. | 4. | 5. | 6. | 7. | 8. |
|---|---|---|---|---|---|---|---|---|
| 1. Perceived Risk | - | | | | | | | |
| 2. Government Trust - Single | -.31* | - | | | | | | |
| 3. Government Trust - Index | -.12* | .64* | - | | | | | |
| 4. Industry Trust | -.39* | .68* | .50* | - | | | | |
| 5. AI Trust | -.43* | .62* | .39* | .67* | - | | | |
| 6. Slowdown Support | .60* | -.28* | -.05 | -.41* | -.46* | - | | |
| 7. Ban Support | .64* | -.20* | -.01 | -.26* | -.38* | .64* | - | |
| 8. Exposure Frequency | -.11* | .42* | .26* | .41* | .49* | -.21* | -.12* | - |
| *M (SD)* | 4.56 (1.62) | 3.09 (1.77) | 4.24 (1.56) | 3.34 (1.83) | 3.91 (1.62) | 4.79 (1.56) | 4.71 (1.63) | 1.38 (1.41) |
| Cronbach's α or *r* | α = .88 | - | α = .64 | α = .76 | α = .90 | α = .77 | α = .79 | *r* = .57 |
| PCF variance proportion | 0.76 | - | 0.59 | 0.69 | 0.77 | 0.7 | 0.71 | 0.78 |
| Principal factor eigenvalue | 2.66 | | 1.07 | 1.45 | 2.7 | 1.70 | 1.6 | 0.90 |
| *Note.* * denotes $p < .001$ | | | | | | | | |

### 3.1.1 DV: Policy Preferences

Rather than using a bipolar measure forcing respondents to select a policy stance, we asked a battery of questions to capture the mixed preferences on the level of support. Two variables are subsequently constructed to demonstrate respondents' policy preferences on softer and stronger regulation strategies. One is the "support for slowing down AI development," which was the average of three items ("I support public campaigns to slow down AI development," "I support government regulation that slows down AI development," and "I oppose government regulation that slows down AI development"), each measured on 1 (strongly disagree) to 7 (strongly agree) Likert-type scales. The opposition to government regulation item was reverse coded prior to inclusion in the composite variable. The other is the "support for banning advanced AI technologies," which was also the average of three items ( "I support banning the development of artificial general intelligence that is smarter than humans," "I support a global ban on data centers that are large enough to train AI systems that are smarter than humans," and "I support a global ban on the development of sentience in robots/AIs") measured on 1 (strongly disagree) to 7 (strongly agree) Likert-type scales. These policy preference items were inspired by Metzinger's (2021) proposal for a moratorium on synthetic phenomenology and popular discourse.

### 3.1.2 IV: Trust

Trust is an umbrella term. For the purpose of this research, we measure and focus on the institutional trust and the AI technology trust. For institutional trust, Williamson (1993) suggested that the institutions of governance indicate how the micro-level structure understands and operates the institutional environment, such as societal culture, politics, regulation, network, bureaucracy, and corporations. Thus, institutional trust could be operationalized as the sensemaking of individuals to the social and organizational contexts  that render their expectations (Dietz, 2011), for example institutional competency (Devos et al., 2002), confidence (Rothstein & Stolle, 2008), and trustworthiness (Grimmelikhuijsen & Knies, 2017; Gulati et al., 2019; Merritt, 2011; Schepman & Rodney, 2022).  In the context of AI development and regulation, we paid specific attention to measuring public opinion regarding their trust in government and AI companies.

#### 3.1.2.1 Trust in Government

Regulatory trust in the government refers to trust in the development and enforcement of regulations (Marusic et al., 2020). Although the common practice is to measure citizens' trust in government using a generic single-item measure, the conceptualization of trust should also trace the causes of trust, such as the perceived competency and ability, the perceived benevolence, and the perceived integrity of the government (Grimmelikhuijsen & Knies, 2017). For this study, the dimension of perceived competency is particularly relevant for measuring trust in the capability and effectiveness of the government to regulate AI technologies and AI development. Therefore, our measure of trust in government is the average of three items that are a proxy for governmental competency to manage and regulate AI ("AI systems include many different parts. To what extent do you trust the following parts…governments?", "To what extent do you agree or disagree that governments have the power to effectively enforce regulations on the development of AI?," and "Do you think that governments have the power to regulate the development of AI?"). The first item was measured on a 1 (not at all) to 7 (very much) sliding scale. The second question was measured on a 1 (strongly disagree) to 7 (strongly agree) Likert-type scale. The third question was measured categorically ("no," "not sure," "yes") and transformed to a numeric scale (no = 1, not sure = 4, yes = 7). Both the principal component variance and principal factor analysis (eigenvalue = 1.07) suggest that the 3-item construct is valid. However, we also provide regression results using the three-item measure and a single-item measure (only the first item) as a robustness check to address any validity concerns that the second and the third items of government trust index do not directly measure citizens' trust, but instead, their perceptions of the government's regulatory competency.

### 3.1.2.2 Trust in AI Industry

Another form of institutional trust is trust in companies developing AI technologies. Based on the basic principle of institutional trust mentioned earlier, we measure citizens' perceptions of AI creators' (i.e., developers) in terms of trust, benevolence, and competency. We averaged three items ("AI systems include many different parts. To what extent do you trust the following parts…companies?", "Do you trust that the creators of large language models (e.g., OpenAI and GPT-4) put safety over profits?", and "Do you trust that the creators of an AI can control all current and future versions of the AI?"). The first item was measured on a 1 (not at all) to 7 (very much) sliding scale. The other two questions were measured categorically ("no," "not sure," "yes") and transformed to numeric scales (no = 1, not sure = 4, yes = 7).

### 3.1.2.3 Trust in AI Technology

Trust in AI is trust of AI technologies. It captures an individual's perceived trust of several popular AI technologies and applications. This measure was the average of four items measured on 1 (strongly disagree) to 7 (strongly agree) Likert-type scales. The stem "I trust" was paired with "chatbots," "large language models," "robots," and "game-playing AI."

### 3.1.3 IV: Perceived Risk

Risk perceptions are considered important in predicting users' willingness to interact with technological artifacts, particularly in the broad technology acceptance theory (Ospina & Pinzón, 2018). Perceived risk was assessed with the average of four items ("AI is likely to cause human extinction," "Robots/AIs may be harmful to me personally," "Robots/AIs may be harmful to people in the USA," and "Robots/AIs may be harmful to future generations of people") measured on 1 (strongly disagree) to 7 (strongly agree) Likert-type scales. The first statement was based on popular discourse in 2023 around extreme risks from AI (e.g., Yudkowsky, 2023), and the latter three were modeled on Thaker et al.'s (2017) perceived risk index in the climate change context.

## 4 Analysis and Results

### 4.1 Analytic Strategy

Based on the research model depicted in Figure 1, we tested the effects of institutional trust, AI trust, and perceived risk on stronger and softer AI regulation preferences. We test the general public's regulatory preferences by categorizing preferences into the stronger approach (ban) and the softer approach (slow down development). It is likely that there would be associations between the residuals of linear regressions predicting softer and stronger regulations that would lead to higher estimation errors given their strong conceptual similarity and the likely strong positive correlation between the outcome variables. Therefore, given the strong positive correlation between the two dependent variables ($r = 0.64$, Table 3), we used seemingly unrelated regressions (SUR) to test our hypotheses. As a special class of linear regression, even for cross-sectional data, SUR involves correlating the residuals of the dependent variables in different regression equations to provide more asymptotically efficient estimates compared to those obtained through single-equation least squares regressions that assume residuals are uncorrelated (Zellner, 1962; King, 1989). The significance of the correlation is assessed by the Breusch-pagan test of independence with the null hypothesis assuming two equations are not related. We detected significant Breusch-pagan tests (see Tables 4a-b), further supporting our use of SURs. Another advantage of using SURs is that they allow an estimate of the between-equation covariances and include postestimation to compare coefficients across equations.

We performed three SUR models and present the standardized beta coefficients and corresponding t-statistics for strength and significance in Table 4a. Each model consisted of two equations with the two dependent variables, support for slowing down AI development and support for banning advanced AI technologies.

Adopting a hierarchical approach, we predicted policy preferences from demographics in step one, Model 1 (baseline): gender (male = 0, female = 1), age, income, race/ethnicity (Asian American = 0), political orientation (very liberal = 1, moderate = 3, very conservative = 5), and familiarity with AI technologies. In step 2, Model 2 (trust model), we added the trust predictor variables: trust in the government, trust in the companies, and trust in AI technologies. In step 3, Model 3 (full model) we added a third step to assess the predictive impact of perceived risk. We check for the multicollinearity problem with the Variance Inflation Factor (VIF). All VIF values are below 5 with a mean VIF of 1.9, suggesting multicollinearity is not a cause for concern.

### 4.2 Regression Findings

The Model 1 results suggest that the effects of demographic variables on support for slowing down AI development and banning the development of advanced AI were small, although several were statistically significant.[1] Older adults were more likely to support both slowing down AI development and banning the development of advanced AI. The effect of age on support for slowing down development was larger than on support for bans. Female respondents more strongly supported both slowing down and banning policies than male respondents. The effect was larger on support for bans than on support for slowing down development. High-income individuals were more likely to support banning the development of advanced AI but there was no effect of income on support for slowing down development. Education level had no impact on policy preferences. Compared with Asian Americans, other racial or ethnic groups showed more support for banning advanced AI technologies but only Black Americans showed more support for slowing down AI development. Conservative political orientation predicted support for bans but there was no effect of political orientation on support for slowing down AI development. Greater self-reported exposure to AI predicted less support for slowing down AI development and did not predict support for bans.

---

[1] We considered coefficients < .20 to be small effects, .20 to .50 to be moderate effects, and > .50 to be large effects.

Model 2 explained significantly more variance in both outcomes than Model 1. Model 1 explained 6-8% of the variance in policy preferences whereas Model 2 explained 20-28% of the variance in policy preferences. This increase in the explanatory power of the model suggests the importance of citizens' trust of entities in the AI ecosystem. Additionally, the effects of the trust variables were moderate in size and larger on support for slowing down development than on support for bans. Specifically, placing more trust in the government significantly predicted more support for slowing down and banning policies. For each standard deviation (SD) increase in government trust, support for slowing down increased by 0.23 SD. For each SD increase in government trust, support for banning policies increased by 0.20 SD. Placing more trust in AI companies and AI technologies significantly predicted less support for both policies. For each SD increase in industry trust, support for slowing down decreased by 0.27 SD and support for bans decreased by .11 SD. For each SD increase in AI trust, slow down support decreased by .37 SD and support for bans decreased by .39 SD. These results highlight the significant role that institutional trust, of the government and companies, plays in influencing policy preferences for AI regulation. The positive predictive impact of trust in the government suggests that people believe that government agencies can offer effective oversight and intervention to ensure the responsible and safe development of AI. In contrast, the negative predictive impact of trust in companies suggests that people who trust AI companies do not believe that government regulation is necessary to slow down or ban certain advanced AI technologies. Those with greater trust in AI companies may have confidence that they will voluntarily implement best practices and safeguards, reducing the need for government intervention.

 Model 3, including perceived risk, again explained significantly more variance in both outcomes than Model 2, increasing from 20-28% variance explained to 46% variance explained. Perceived risk moderately to strongly, positively predicted support for slowing down AI development and banning advanced AI technologies. For each SD increase in perceived risk, support for slowing down AI increased by .49 SD and support for bans increased by .59. These effects were moderate to large while controlling for the effects of the demographic and trust predictor variables, suggesting the relative importance of perceived risk in AI policy preferences. The predictive impact of the trust variables largely remained, although the effects became smaller, except for the disappearance of the significant effect of trust in AI companies on support for bans. These results suggest that Americans are concerned about the potential of AI to harm themselves and all of humanity, and this concern predicts a preference for slowing down and banning the development of advanced AI technologies.

**Table 4a. Predicting Policy Preferences Using SUR Models.**

| | Model 1 Baseline | | Model 2 Trust model | | Model 3 Full model | |
|---|---|---|---|---|---|---|
| | Slowdown | Ban | Slowdown | Ban | Slowdown | Ban |
| **Government Trust - index** | | | 0.23*** (7.32) | 0.20*** (6.02) | 0.16*** (5.53) | 0.12*** (4.19) |
| **Industry Trust** | | | -0.27*** (-6.82) | -0.11** (-2.60) | -0.14*** (-4.14) | 0.05 (1.38) |
| **AI Trust** | | | -0.37*** (-9.67) | -0.39*** (-9.35) | -0.18*** (-4.88) | -0.16*** (-4.32) |
| **Perceived Risk** | | | | | 0.49*** (18.37) | 0.59*** (21.92) |
| **Control variables** | | | | | | |
| **Age** | 0.10** (3.17) | 0.07* (2.15) | 0.03 (1.13) | 0.01 (0.41) | 0.09*** (3.59) | 0.08*** (3.50) |
| **Female** | 0.14*** (4.06) | 0.17*** (5.00) | 0.07* (2.26) | 0.11*** (3.63) | 0.06* (2.42) | 0.11*** (3.96) |
| **Income (per 1K)** | 0.03 (0.86) | 0.07* (2.00) | 0.06 (1.63) | 0.09** (2.58) | 0.03 (1.05) | 0.06 (1.92) |
| **Education** | 0.03 (0.88) | -0.00 (-0.02) | 0.02 (0.61) | -0.00 (-0.14) | 0.03 (0.95) | 0.01 (0.17) |
| **Black American** | 0.10* (2.22) | 0.11* (2.48) | 0.05 (1.32) | 0.07 (1.56) | 0.05 (1.65) | 0.07* (1.96) |
| **Hispanic American** | 0.07 (1.48) | 0.11* (2.21) | 0.02 (0.44) | 0.06 (1.16) | 0.02 (0.44) | 0.06 (1.41) |
| **Indigenous** | 0.04 (1.16) | 0.07* (2.09) | 0.01 (0.31) | 0.05 (1.92) | -0.02 (-0.96) | 0.02 (0.60) |
| **White American** | 0.11 (1.78) | 0.14* (2.32) | 0.04 (0.77) | 0.08 (1.37) | -0.03 (-1.27) | 0.08 (1.60) |
| **Other Ethnicity** | 0.07 (1.83) | 0.08* (2.41) | 0.01 (0.35) | 0.04 (1.28) | 0.03 (0.75) | 0.06* (2.38) |
| **Political Orientation** | 0.02 (0.63) | 0.11*** (3.21) | 0.01 (0.44) | 0.10** (3.15) | -0.03 (-0.95) | 0.04 (1.73) |
| **Exposure Frequency** | -0.16*** (-4.68) | -0.06 (-1.76) | 0.04 (1.11) | 0.09* (2.55) | 0.16*** (5.53) | 0.01 (0.51) |
| **Model Information** | | | | | | |
| **Observation** | 1098 | | 1098 | | 1098 | |
| **R²** | 0.08 | 0.06 | 0.28 | 0.20 | 0.46 | 0.46 |
| **Correlation(ϵϵ′)** | 0.62 | | 0.547 | | 0.372 | |
| **Breusch-pagan test** | $X^2$=373.1, $p$<.000 | | $X^2$=175.6, $p$<.000 | | $X^2$=165.2, $p$<.000 | |

*Note.* * $p$ <0.05, ** $p$ <0.01, and *** $p$ <0.001. The race/ethnicity reference group is Asian Americans. Standardized beta coefficients; t-statistics in parentheses.

For a robustness check to address the potential that the index measure of government trust assesses more than trust (e.g., competency, power), we re-ran the SUR models using only the single-item that explicitly measures government trust (i.e., extent of trust in the government element of the AI ecosystem; Table 4b). Compared with the models using the index measure (Table 4a), the effect size of the single-item government trust measure is reduced, but the coefficient direction and significance remains consistent. This indicates the potential importance of addressing citizens' trust of the government including perceptions of government competency and power, and knowledge of AI governance in practice. Another noteworthy difference between the original and robust models is the increased effect size of perceived risk in the models with the single-item government trust measures (Table 4b). This suggests some positive covariance between the index measure of government trust and citizens' perceived risk.

**Table 4b. Predicting Policy preferences using SUR models (a robustness check using the single-item government trust).**

| | Model 4 Trust model | | Model 5 Full model | |
|---|---|---|---|---|
| | Slowdown | Ban | Slowdown | Ban |
| **Government Trust - single** | 0.09* (2.30) | 0.09* (2.17) | 0.08* (2.40) | 0.08* (2.34) |
| **Industry Trust** | -0.21*** (-5.29) | -0.07 (-1.60) | -0.10** (-2.94) | 0.06 (1.82) |
| **AI Trust** | -0.38*** (-9.54) | -0.40*** (-9.57) | -0.18*** (-5.10) | -0.16*** (-4.60) |
| **Perceived Risk** | | | 0.51*** (19.41) | 0.61*** (23.32) |
| **Control variables** | | | | |
| **Age** | 0.02 (0.76) | 0.00 (0.10) | 0.08*** (3.31) | 0.08*** (3.04) |
| **Female** | 0.07* (2.38) | 0.11*** (3.61) | 0.07** (2.58) | 0.11*** (4.20) |
| **Income (per 1K)** | 0.05 (1.64) | 0.09* (2.53) | 0.03 (0.98) | 0.06* (1.99) |
| **Education** | 0.04 (1.12) | 0.01 (0.27) | 0.04 (1.37) | 0.01 (0.41) |
| **Black American** | 0.07 (1.67) | 0.09* (1.97) | 0.07 (1.82) | 0.08* (2.27) |
| **Hispanic American** | 0.04 (0.76) | 0.07 (1.49) | 0.03 (0.66) | 0.06 (1.56) |
| **Indigenous** | 0.01 (0.31) | 0.05 (1.82) | -0.02 (-0.98) | 0.02 (0.62) |
| **White American** | 0.05 (0.87) | 0.09 (1.58) | 0.04 (0.85) | 0.09 (1.74) |
| **Other Ethnicity** | 0.01 (0.28) | 0.04 (1.12) | 0.02 (0.84) | 0.06* (1.99) |
| **Political Orientation** | 0.00 (0.03) | 0.09** (3.09) | -0.04 (-1.75) | 0.04 (1.63) |
| **Exposure Frequency** | 0.03 (0.97) | 0.09* (2.51) | -0.03 (-1.19) | 0.01 (0.28) |
| **Model Information** | | | | |
| **Observation** | 1098 | | 1098 | |
| **R²** | 0.25 | 0.17 | 0.44 | 0.45 |
| **Correlation(ϵϵ′)** | 0.563 | | 0.383 | |
| **Breusch-pagan test** | $X^2$=348.8, $p$<.000 | | $X^2$=161.8, $p$<.000 | |

*Note.* * $p<0.05$, ** $p<0.01$, and *** $p<0.001$. The race/ethnicity reference group is Asian Americans. Standardized beta coefficients; t-statistics in parentheses.

# 5 General Discussion

## 5.1 Results Interpretation

Support for slowing down AI and support for banning certain AI technologies were strongly related to each other and both were predicted moderately to strongly by trust and perceived risk. These moderate to strong effects in this large, nationally representative sample provide empirical evidence for the importance of understanding US citizens' trust in different aspects of the AI ecosystem and the risks they perceive from AI. For context, recent studies of the effect of trust in the government and perceived risk on responses to the COVID-19 pandemic have found small to moderate effects of these constructs on preventive health behavior and compliance with precautionary measures (Bruine de Bruin et al., 2020; Han et al., 2023; Shanka & Menebo, 2022; Trent et al., 2022). The research on trust and risk predictors of AI policy preferences is sparse, however some market research has shown small effects of perceived risk and trust in AI voice assistants on brand loyalty (Hasan et al., 2021). The moderate to large effects we observed in this analysis of the AIMS survey data surpass these previously documented results and suggest that trust and risk constructs are critical to AI policy preferences.

Institutional trust was associated with policy support, but, the institution type, whether governmental or industrial, shifted the direction of the effect. Trusting the government to regulate AI development was associated with more support for regulation policies, indicating some belief in the efficacy of governments. Individuals could place higher trust in government institutions to either slow down or prohibit AI development in order to protect the public interest and address potential AI-related risks, which could be an effective means to mitigate potential harm. Trusting companies to safely develop AI technologies was associated with less support for regulation policies, suggesting that there may be some belief in the U.S. that legislative governance may not be necessary if AI companies are trustworthy. The negative relationship between trust in private companies and support for slowdowns in AI development may be driven by the belief that industry self-regulation and innovation are sufficient to address any challenges posed by AI technologies.

In the AI context, where corporate and government stakeholders have potentially divergent motivations and behaviors, their effects are empirically and conceptually separable. By assessing the effects of public trust in these two powerful institutions, we obtained a more meaningful and nuanced understanding of the spectrum of institutional trust effects in AI governance. In addition to these constructs showing differing effect directions in the SUR models, their correlation ($r = .50$) suggested that they are different, but related, forms of trust. A strong, positive correlation of .90 or higher (Rönkkö & Cho, 2022) would have indicated that our choice to separate institutional trust into government trust and industry trust was invalid. Further, both government and industry trust were moderately to strongly positively correlated with trust of AI technologies ($r = .39, .67$), but again, not over the .90 threshold that would indicate that these concepts are indistinguishable. The empirical evidence suggested government trust, industry trust, and trust in AI technologies are meaningfully distinct. Further, the similarity between the model results for the composite index and single-item measures of government trust increase our credence in the robustness of these effects and suggest that further targeted study of government trust in the AI context is needed.

The results support our hypotheses and map onto intuitive expectations for these relationships given popular discourse (see Table 5). Explanations for these results deserve more attention in future research. In our study, the perceived risk construct mostly concerned perceived safety. The association between heightened perceived risk resulting from a safety concern and increased desire to regulate is intuitively comprehensible given humans evolutionarily-based desire to survive (Schaller et al., 2017) and the connection of perceived risk to policy support in other domains that have safety implications such as public health (Buykx et al., 2015), climate change (Leiserowitz, 2006), and nuclear power (Stoutenborough et al., 2013). Americans worry that AI could make critical errors and harm individuals. This concern predicts regulatory support.

In addition to the effects of trust and risk, several demographic variables predicted policy support, although the same pattern of participant characteristics did not predict both support for slowdown and support for bans. Support for slowing down development was predicted by being older, being female, being Black compared to Asian, and having less exposure (i.e., familiarity) to AI. Support for bans was predicted by being older, being female, having a higher income, being Black, Hispanic, Indigenous, Other, or White compared to Asian, and having a more conservative political orientation. Education level was not predictive of either regulation policies. The effects of income might be caused by high income individuals' worries about a more powerful social artifact and its societal consequences.

**Table 5. Summary of Results by Hypothesis.**

| Hypothesis | Result |
|---|---|
| Hypothesis 1a: Trust in the government increases the support of softer AI regulation. | Supported |
| Hypothesis 1b: Trust in the government increases the support of stronger AI regulation. | Supported |
| Hypothesis 2a: Trust in the company decreases the support of softer AI regulation. | Supported |
| Hypothesis 2b: Trust in the company decreases the support of stronger AI regulation. | Partially supported |
| Hypothesis 3a: Trust in the AI decreases the support of softer AI regulation. | Supported |
| Hypothesis 3b: Trust in the AI decreases the support of stronger AI regulation. | Supported |
| Hypothesis 4a: Perceived risks increase the support of softer AI regulation. | Supported |
| Hypothesis 4b: Perceived risks increase the support of stronger AI regulation. | Supported |

## 5.2 Results Implication

Our research results have several theoretical implications, in particular for expanding the existing research on AI regulatory preferences. The existing research analyzes the public attitudes toward AI regulation and the determinants of regulatory support (Heinrich & Witko, 2024;

O'Shaughnessy et al., 2023). We identified factors relevant to citizens' preferences for AI regulation to deepen our understanding of this issue. First, citizens' trust in the government was associated with higher support of regulatory policies, regardless of regulation strength. The relationship between government trust and support of AI regulatory policy is consistent with the existing research from other policy contexts (Harring & Jagers, 2013; Huber & Wicki, 2021; Kallbekken & Sælen, 2011; Kulin & Johansson Sevä, 2021). Also, the results indicate that government trust is associated with higher support for both softer and stronger regulatory policy preferences. In other words, when citizens are confident in the public sector, they tend to support regulation.

Second, the effects of trust in industry on policy preferences depended on whether risk is also included in the model. In most of our models, industry trust was associated with a decrease in citizens' support of both softer and stronger AI regulations. The results are similar to the existing body of knowledge that citizens with more trust in the private sector prefer the government not to intervene in market operation and development (Cvetkovich et al., 2002; Edlund & Lindh, 2013; Rhodes et al., 2017). However, when the model includes perceived AI risk, the impact of trust in the company on the support of stronger AI regulation becomes insignificant, indicating trust in the private sector may be most important for its effect on risk perception.

Third, high trust in AI may reduce citizens' support for the government's intervention in technology development and applications. This result is consistent with existing studies on environmental regulations and AI acceptance (Ahn & Chen, 2022; Davidovic et al., 2020; Fairbrother, 2016; Kulin & Johansson Sevä, 2021). Our findings indicated that citizens' trust in AI facilitates their preference for innovating and deploying this technology, which is in line with existing digital governance research (Horsburgh et al., 2011; Lin et al., 2021; Song & Lee, 2016; Tolbert & Mossberger, 2006; Wang et al., 2023).

### 5.3 Limitations and Directions for Future Research

The AIMS supplemental survey data covered a range of novel topics on which public opinion may quickly evolve in the coming years as there are new government policies, new ways of using AI in society, and new forms of AI technology itself (Anthis et al., 2025; Pauketat et al., 2025). In light of the uncertainty witnessed across questions (e.g., "not sure"), and because the particular context of our survey questions—focused on AI risks and safety—participants may have been steered in particular directions. These limitations were addressed in part by differently worded questions within the survey to interrogate the same underlying attitudes and beliefs, but it is not clear that the conceptualizations we generate for the current study, such as the three types of trust, would persist in other examinations. In particular, the government trust measure we constructed might serve as a good proxy to the government's competency in enforcing the specific AI regulation, but might be less applicable to a broader scope of citizens' trust in the government, meaning we caution against overinterpretation of these results.

Our survey inevitably has another limitation, the risk of common source bias due to the nature of cross-sectional surveys and measuring the key variables in a single survey (Favero & Bullock, 2015). However, as suggested by Favero and Bullock, the only remedy for common source bias is to rely on an independent source of data through deliberate survey design. The AIMS survey was designed carefully and the details can be openly accessed on the pre-registry OSF site (https://osf.io/7p2wt/).

Our focus on the U.S. context provides a sense of public opinion among residents of a major, if not the most powerful, country in the rapidly developing field of AI. However, given the numerous different cultural associations with AI and its effects on different domains of social life across populations, it is likely that public opinion varies widely across countries. It is possible that regions with more demographic and cultural similarity to the U.S. have more similar public opinion, and future research should explore this variation. Further, although we estimated and controlled for racial/ethnic group-based and political orientation effects in the SUR models, there were some small effects that may deserve attention in future research. The broadly generalizable nationally representative results we presented cannot adequately address the specific and nuanced views of AI, trust, and perceived risk based on racial/ethnic or political identities.

In general, our aim was to gather robust data to create a baseline of public opinion data for future work, including experimental research that teases out the causal relationships undergirding these constructs—which we were not able to test with this data—as well as repeated survey measures to track the dynamic of public attitudes over time and measures of real-world behavior in addition to public opinion. We are particularly eager to see similar data collected outside of the U.S. with which we can understand cultural variation and global dynamics as this nascent generative AI technology, or "digital minds," begins to reshape governance and public life.

## 6 Conclusion

Rapid AI-based technological change has presented many opportunities, and significant societal risks. Although governments have begun to introduce legislation to regulate AI developments, public opinion on the severity of risks and trust in the AI ecosystem has been neglected despite its role in shaping governance. This study responded to calls for empirical data and conceptual development to make sense of how public opinion may fuel and steer the effects of generative AI and emergent digital minds on AI governance and public administration. In particular, we built on conceptualizations of institutional and AI trust to examine the predictive effects of trust in various parts of the AI ecosystem alongside perceptions of AI risks on support for soft and strong regulatory policies.

With a unique U.S. nationally representative survey leveraging data from 2023, we found varying levels of trust in governments to regulate AI developments, companies to control AI developments, and AI technologies themselves, and a high degree of AI risk perception. Support for institutional policies to slow down or ban AI developments was strong. The predictive models supported our conceptualization that institutional trust in governments and industries, AI trust, and risk perceptions matter for good AI governance. Increased trust in the regulatory power of governments was important for regulatory preferences, suggesting support for and a belief in the effectiveness of government regulation in the AI context. Decreased institutional trust in AI companies and trust in AI technologies predicted increased support for regulation, meaning that people may rely on government-based regulation to bound fast-paced technological changes that they are not prepared for.

The U.S. public supports safety regulations such as slowing down development and banning the development of some systems, and this support depends on trust and risk perceptions. The nationally representative Artificial Intelligence, Morality, and Sentience (AIMS) data suggest that AI safety and greater regulation of AI developments is desired for good governance of AI, perhaps in contrast to the current accelerationist policies of U.S. policymakers. Overall, these data and our analyses suggest a complex tapestry of trust and risk perceptions, and point to the importance of disentangling evaluations of governments, companies, and AI technologies. As societal risks and digital minds continue to emerge from ongoing AI innovation, we encourage further research into the understudied dynamics of trust in various parts of the AI ecosystem, risk perceptions, and AI governance. For researchers and policymakers, we urge a sense of mindfulness to the opportunities of AI alongside caution to mitigate the widely perceived risks of AI as technological advances continue to shape societal risk perception and public opinion on AI governance.

# References

Ahn, M. J., & Chen, Y.-C. (2022). Digital transformation toward AI-augmented public administration: The perception of government employees and the willingness to use AI in government. *Government Information Quarterly*, 39(2), 101664. https://doi.org/10.1016/j.giq.2021.101664

Androutsopoulou, A., Karacapilidis, N., Loukis, E., & Charalabidis, Y. (2019). Transforming the communication between citizens and government through AI-guided chatbots. *Government Information Quarterly*, 36(2), 358-367. https://doi.org/10.1016/j.giq.2018.10.001

Anthis, J. R. (2023, March 6). Key questions for digital minds. *Sentience Institute*. https://www.sentienceinstitute.org/blog/key-questions-for-digital-minds

Anthis, J. R., Pauketat, J. V. T., Ladak, A., & Manoli, A. (2025). *Perceptions of sentient AI and other digital minds: Evidence from the AI, Morality, and Sentience (AIMS) survey*. arXiv. https://doi.org/10.48550/arXiv.2407.08867

Aoki, N. (2020). An experimental study of public trust in AI chatbots in the public sector. *Government Information Quarterly*, 37(4), 101490. https://doi.org/10.1016/j.giq.2020.101490

Barkane, I. (2022). Questioning the EU proposal for an Artificial Intelligence Act: The need for prohibitions and a stricter approach to biometric surveillance. *Information Polity*, 27(2), 147-162. https://doi.org/10.3233/IP-211524

Bode, I., & Huelss, H. (2023). Constructing expertise: the front-and back-door regulation of AI's military applications in the European Union. *Journal of European Public Policy*, 30(7), 1230-1254. https://doi.org/10.1080/13501763.2023.2174169

Bruine de Bruin, W., Saw, H-W., Goldman, D. P. (2020). Political polarization in US residents' COVID-19 risk perceptions, policy preferences, and protective behaviors. *Journal of Risk and Uncertainty*, 61, 177-194. https://doi.org/10.1007/s11166-020-09336-3

Bullock, J. (2019). Artificial intelligence, discretion, and bureaucracy. *The American Review of Public Administration*, 49(7), 751-761. https://doi.org/10.1177/0275074019856123

Bullock, J. B., Chen, Y.-C., Himmelreich, J., Hudson, V. M., Korinek, A., Young, M. M., & Zhang, B. (2023). Introduction. In J. B. Bullock, Y.-C. Chen, J. Himmelreich, V. M. Hudson, A. Korinek, M. M. Young, & B. Zhang (Eds.), *The Oxford Handbook of AI Governance* (pp. 0). Oxford University Press. https://doi.org/10.1093/oxfordhb/9780197579329.013.1

Bullock, J. B., Greer, R. A., & O'Toole Jr, L. J. (2019). Managing risks in public organizations: A conceptual foundation and research agenda. *Perspectives on Public Management and Governance*, 2(1), 75-87. https://doi.org/10.1093/ppmgov/gvx016

Buykx, P., Gilligan, C., Ward, B., Kippen, R., & Chapman, K. (2015). Public support for alcohol policies associated with knowledge of cancer risk. *International Journal of Drug Policy*, 26(4), 371-379. https://doi.org/10.1016/j.drugpo.2014.08.006

Cath, C. (2018). Governing artificial intelligence: ethical, legal and technical opportunities and challenges. *Philosophical Transactions of the Royal Society A: Mathematical, Physical and Engineering Sciences*, 376(2133), 20180080. https://doi.org/10.1098/rsta.2018.0080

Chen, Y.-C., Ahn, M., & Wang, Y.-F. (2023). Artificial Intelligence and public values: Value impacts and governance in the public sector. *Sustainability*, 15(6), 4796. https://doi.org/10.3390/su15064796

Chen, T., & Gasco-Hernandez, M. (2024). Uncovering the results of AI Chatbot use in the public sector: Evidence from US state governments. *Public Performance & Management Review*, 1-26. https://www.tandfonline.com/doi/full/10.1080/15309576.2024.2389864

Coeckelbergh, M. (2022). *The political philosophy of AI: An introduction*. John Wiley & Sons.

Cvetkovich, G., Siegrist, M., Murray, R., & Tragesser, S. (2002). New information and social trust: Asymmetry and perseverance of attributions about hazard managers. *Risk Analysis*, 22(2), 359-367. https://doi.org/10.1111/0272-4332.00030

Davidovic, D., Harring, N., & Jagers, S. C. (2020). The contingent effects of environmental concern and ideology: institutional context and people's willingness to pay environmental taxes. *Environmental Politics*, 29(4), 674-696. https://doi.org/10.1080/09644016.2019.1606882

De Hert, P., & Bouchagiar, G. (2022). Visual and biometric surveillance in the EU. Saying 'no'to mass surveillance practices? *Information Polity*, 27(2), 193-217. https://doi.org/10.3233/IP-211525

Devos, T., Spini, D., & Schwartz, S. H. (2002). Conflicts among human values and trust in institutions. *British Journal of Social Psychology*, 41, 481-494. https://psycnet.apa.org/doi/10.1348/014466602321149849

Dietz, G. (2011). Going back to the source: Why do people trust each other? *Journal of Trust Research*, 1(2), 215-222. https://doi.org/10.1080/21515581.2011.603514

Dietz, T., Dan, A., & Shwom, R. (2007). Support for climate change policy: Social psychological and social structural influences. *Rural sociology*, *72*(2), 185-214. https://doi.org/10.1526/003601107781170026

Edlund, J., & Lindh, A. (2013). Institutional trust and welfare state support: on the role of trust in market institutions. *Journal of Public Policy*, 33(3), 295-317. https://doi.org/10.1017/S0143814X13000160

Fairbrother, M. (2016). Trust and public support for environmental protection in diverse national contexts. *Sociological Science*, 3, 359-382. https://doi.org/10.15195/v3.a17

Fairbrother, M., Sevä, I. J., & Kulin, J. (2019). Political trust and the relationship between climate change beliefs and support for fossil fuel taxes: Evidence from a survey of 23 European countries. *Global Environmental Change*, 59, 102003. https://doi.org/10.1016/j.gloenvcha.2019.102003

Favero, N., & Bullock, J. B. (2015). How (not) to solve the problem: An evaluation of scholarly responses to common source bias. *Journal of Public Administration Research and Theory*, 25(1), 285-308. https://doi.org/10.1093/jopart/muu020

Fountain, J. E. (2022). The moon, the ghetto and artificial intelligence: Reducing systemic racism in computational algorithms. *Government Information Quarterly*, 39(2), 101645. https://doi.org/10.1016/j.giq.2021.101645

Fui-Hoon Nah, F., Zheng, R., Cai, J., Siau, K., & Chen, L. (2023). Generative AI and ChatGPT: Applications, challenges, and AI-human collaboration. *Journal of Information Technology Case and Application Research*, 25(3), 277-304. https://doi.org/10.1080/15228053.2023.2233814

Gesk, T. S., & Leyer, M. (2022). Artificial intelligence in public services: When and why citizens accept its usage. *Government Information Quarterly*, 39(3), 101704. https://doi.org/10.1016/j.giq.2022.101704

Grimmelikhuijsen, S. (2023). Explaining why the computer says no: Algorithmic transparency affects the perceived trustworthiness of automated decision-making. *Public Administration Review*, 83, 241-262. https://doi.org/10.1111/puar.13483

Grimmelikhuijsen, S., & Knies, E. (2017). Validating a scale for citizen trust in government organizations. *International Review of Administrative Sciences*, 83(3), 583-601. https://doi.org/10.1177/0020852315585950

Grimmelikhuijsen, S., & Meijer, A. (2022). Legitimacy of Algorithmic Decision-Making: Six Threats and the Need for a Calibrated Institutional Response. *Perspectives on Public Management and Governance*, 5(3), 232-242. https://doi.org/10.1093/ppmgov/gvac008

Gulati, S., Sousa, S., & Lamas, D. (2019). Design, development and evaluation of a human-computer trust scale. Behaviour & Information Technology, 1004-1015. https://doi.org/10.1080/0144929X.2019.1656779

Han, Q., Zheng, B., Cristea, M., Agostini, M., Bélanger, J. J., Gützkow, B., Kreienkamp, J., Leander, N. P., PsyCorona Collaboration (2023). Trust in government regarding COVID-19 and its associations with preventive health behaviour and prosocial behaviour during the pandemic: A cross-sectional and longitudinal study. *Psychological Medicine*, 53(1), 149-159. https://doi.org/10.1017/S0033291721001306

Harring, N. (2018). Trust and state intervention: Results from a Swedish survey on environmental policy support. *Environmental Science & Policy*, *82*, 1-8. https://doi.org/10.1016/j.envsci.2018.01.002

Harring, N., & Jagers, S. C. (2013). Should we trust in values? Explaining public support for pro-environmental taxes. *Sustainability*, 5(1), 210-227. https://doi.org/10.3390/su5010210

Hasan, R., Shams, R., & Rahman, M. (2021). Consumer trust and perceived risk for vocie-controlled artificial intelligence: The case of Siri. *Journal of Business Research*, 131, 591-597. https://doi.org/10.1016/j.jbusres.2020.12.012

Heinrich, T., & Witko, C. (2024). Self-interest and preferences for the regulation of artificial intelligence. *Journal of Information Technology & Politics*, 1-16. https://doi.org/10.1080/19331681.2024.2370815

Ho, S. S., & Chuah, A. S. (2021). Why support nuclear energy? The roles of citizen knowledge, trust, media use, and perceptions across five Southeast Asian countries. *Energy Research & Social Science*, *79*, 102155. https://doi.org/10.1016/j.erss.2021.102155

Horsburgh, S., Goldfinch, S., & Gauld, R. (2011). Is public trust in government associated with trust in e-government? *Social Science Computer Review*, 29(2), 232-241. https://doi.org/10.1177/0894439310368130

Huang, H., Kim, K.-C., Young, M. M., & Bullock, J. B. (2022). A matter of perspective: Differential evaluations of artificial intelligence between managers and staff in an experimental simulation. *Asia Pacific Journal of Public Administration*, 44(1), 47-65. https://doi.org/10.1080/23276665.2021.1945468

Huber, R. A., & Wicki, M. (2021). What explains citizen support for transport policy? the roles of policy design, trust in government and proximity among Swiss citizens. *Energy Research & Social Science*, 75, 101973. https://doi.org/10.1016/j.erss.2021.101973

Ingrams, A., Kaufmann, W., & Jacobs, D. (2022). In AI we trust? Citizen perceptions of AI in government decision making. *Policy & Internet,* 14(2), 390-409. https://doi.org/10.1002/poi3.276

Jung, H., & Camarena, L. (2024). Street-level bureaucrats & AI interactions in public organizations: An identity based framework. *Public Performance & Management Review*, 1-30. https://doi.org/10.1080/15309576.2024.2447352

Kallbekken, S., & Sælen, H. (2011). Public acceptance for environmental taxes: Self-interest, environmental and distributional concerns. *Energy Policy*, 39(5), 2966-2973. https://doi.org/10.1016/j.enpol.2011.03.006

King, G. (1989). A Seemingly Unrelated Poisson Regression Model. *Sociological Methods & Research*, 17(3), 235-255. https://doi.org/10.1177/0049124189017003001

Kitt, S., Axsen, J., Long, Z., & Rhodes, E. (2021). The role of trust in citizen acceptance of climate policy: Comparing perceptions of government competence, integrity and value similarity. *Ecological Economics*, *183*, 106958. https://doi.org/10.1016/j.ecolecon.2021.106958

Korzynski, P., Mazurek, G., Altmann, A., Ejdys, J., Kazlauskaite, R., Paliszkiewicz, J., Wach, K., & Ziemba, E. (2023). Generative artificial intelligence as a new context for management theories: analysis of ChatGPT. *Central European Management Journal*, 31(1), 3-13. https://doi.org/10.1108/CEMJ-02-2023-0091

Kulin, J., & Johansson Sevä, I. (2021). Who do you trust? How trust in partial and impartial government institutions influences climate policy attitudes. *Climate Policy,* 21(1), 33-46. https://doi.org/10.1080/14693062.2020.1792822

Ladak, A., Loughnan, S., & Wilks, M. (2023). The moral psychology of artificial intelligence. *Current Directions in Psychological Science*, 09637214231205866. https://doi.org/10.1177/09637214231205866

Laux, J., Wachter, S., & Mittelstadt, B. (2023). Trustworthy artificial intelligence and the European Union AI act: On the conflation of trustworthiness and acceptability of risk. *Regulation & Governance*, 18(1), 3-32. https://doi.org/10.1111/rego.12512

Leiserowitz, A. (2006). Climate change risk perception and policy preferences: The role of affect, imagery, and values. *Climatic Change*, 77, 45-72. https://doi.org/10.1007/s10584-006-9059-9

Lin, J., Carter, L., & Liu, D. (2021). Privacy concerns and digital government: exploring citizen willingness to adopt the COVIDSafe app. *European Journal of Information Systems*, 30(4), 389-402. https://doi.org/10.1080/0960085X.2021.1920857

Manoli, A., Pauketat, J. V. T., & Anthis, J. R. (2025). *The AI Double Standard: Humans Judge All AIs for the Actions of One*. arXiv. https://doi.org/10.48550/arXiv.2412.06040

Marusic, A., Ghossein, T., Dalton, A. G., & Nielsen, W. (2020). Trust between public and private sectors: The path to better regulatory compliance?. in (Vol. 1 of 2): *A review of the State of Trust, its Importance in Public and Private Sector Relationships, and Policy Implications from Behavioral Insights in Trust -Building Reforms* (English). Washington, D.C. : World Bank Group. http://documents.worldbank.org/curated/en/251541607315661291

Merritt, S. M. (2011). Affective processes in human-automation interactions. *Human Factors,* 53(4), 356-370. https://doi.org/10.1177/0018720811411912

Metzinger, T. (2021). Artificial suffering: An argument for a global moratorium on synthetic phenomenology. *Journal of Artificial Intelligence and Consciousness*, 8(1). 1-24. https://doi.org/10.1142/S270507852150003X

Mei, H., & Zheng, Y. (2024). How M-government services build relative trust? The mediating roles of value creation and risk perception. *Public Performance & Management Review*, *47*(6), 1327-1355. https://doi.org/10.1080/15309576.2024.2370935

Moon, M. J. (2023). Searching for Inclusive Artificial Intelligence for Social Good: Participatory Governance and Policy Recommendations for Making AI More Inclusive and Benign for Society. *Public Administration Review*, 83(6), 1496-1505. https://doi.org/10.1111/puar.13648

Nam-Speers, J., Berry, F. S., & Choi, D. (2023). Examining the role of perceived risk and benefit, shared concern for nuclear stigmatization, and trust in governments in shaping citizen risk acceptability of a nuclear power plant. *The Social Science Journal*, *60*(4), 695-714. https://doi.org/10.1080/03623319.2020.1750846

Netherlands Court of Audit. (2024). Focus on AI in the Dutch central government, https://english.rekenkamer.nl/binaries/rekenkamer-english/documenten/reports/2024/10/16/focus-on-ai-in-central-government/Focus+on+AI+in+the+Dutch+central+government.pdf

Nieuwenhuizen, E. N., Meijer, A. J., Bex, F. J., & Grimmelikhuijsen, S. G. (2024). Explanations increase citizen trust in police algorithmic recommender systems: findings from two experimental tests. *Public Performance & Management Review*, 1-36. https://doi.org/10.1080/15309576.2024.2443140

Ospina, M. L. C., & Pinzón, B. H. D. (2018). Theoretical perspectives on usage of e-government services: A literature review. In *Twenty-fourth American Conference on Information Systems.* https://aisel.aisnet.org/amcis2018/eGovernment/Presentations/7

O'Shaughnessy, M. R., Schiff, D. S., Varshney, L. R., Rozell, C. J., & Davenport, M. A. (2023). What governs attitudes toward artificial intelligence adoption and governance? *Science and Public Policy*, *50*(2), 161-176. https://doi.org/10.1093/scipol/scac056

Pauketat, J. V. T., Bullock, J., & Anthis, J. R. (2023). *Public Opinion on AI Safety: AIMS 2023 Supplement*. PsyArXiv. https://doi.org/10.31234/osf.io/jv9rz

Pauketat, J.V. T., Ladak, A., & Anthis, J.R. (2023). *Artificial Intelligence, Morality, and Sentience (AIMS) Survey* (Mendeley Data; Version V2) [data set]. https://doi.org/10.17632/x5689yhv2n.2

Pauketat, J. V. T., Ladak, A., & Anthis, J. R. (2025). World-making for a future with sentient AI. *British Journal of Social Psychology*, *64*(1), e12844. https://doi.org/10.1111/bjso.12844

Rhodes, E., Axsen, J., & Jaccard, M. (2017). Exploring citizen support for different types of climate policy. *Ecological Economics*, 137, 56-69. https://doi.org/10.1016/j.ecolecon.2017.02.027

Rönkkö, M., & Cho, E. (2022). An updated guideline for assessing discriminant validity. Organizational Research Methods, 25(1), 6-14. https://doi.org/10.1177/1094428120968614

Rothstein, B. & Stolle, D. (2008). The state and social capital: An institutional theory of generalized trust. *Comparative Politics*, 40(4), 441-459. https://www.jstor.org/stable/20434095

Salah, M., Abdelfattah, F., & Al Halbusi, H. (2023). Generative Artificial Intelligence (ChatGPT & Bard) in public administration research: A double-edged sword for street-level bureaucracy studies. *International Journal of Public Administration*, 1-7. https://doi.org/10.1080/01900692.2023.2274801

Saura, J. R., Ribeiro-Soriano, D., & Palacios-Marqués, D. (2022). Assessing behavioral data science privacy issues in government artificial intelligence deployment. *Government Information Quarterly*, 39(4), 101679. https://doi.org/10.1016/j.giq.2022.101679

Schaller, M., Kenrick, D. T., Neel, R., Neubery, S. L. (2017). Evolution and human motivation: A fundamental motives framework. *Social and Personality Psychology Compass*, 11(6), e12319. https://doi.org/10.1111/spc3.12319

Schepman, A., & Rodway, P. (2022). The general attitudes towards Artificial Intelligence scale (GAAIS): Confirmatory validation and associations with personality, corporate distrust, and general trust. *International Journal of Human-Computer Interaction*, 39(13), 2724-2741. https://doi.org/10.1080/10447318.2022.2085400

Schmidt, A. M., Kowitt, S. D., Myers, A. E., & Goldstein, A. O. (2018). Attitudes towards potential new tobacco control regulations among US adults. *International journal of environmental research and public health*, *15*(1), 72. https://doi.org/10.3390/ijerph15010072

Shanka, M. S., & Menebo, M. M. (2022). When and how trust in government leads to compliance with COVID-19 precautionary measures. *Journal of Business Research*, 139, 1275-1283. https://doi.org/10.1016/j.jbusres.2021.10.036

Six, F., & Verhoest, K. (2017). *Trust in regulatory regimes: scoping the field* (pp. 1-36). Edward Elgar Publishing.

Song, C., & Lee, J. (2016). Citizens' use of social media in government, perceived transparency, and trust in government. *Public Performance & Management Review*, 39(2), 430-453. https://doi.org/10.1080/15309576.2015.1108798

Stoutenborough, J. W., Sturgess, S. G., & Vedlitz, A. (2013). Knowledge, risk, and policy support: Public perceptions of nuclear power. *Energy Policy*, *62*, 176-184. https://doi.org/10.1016/j.enpol.2013.06.098

Tan, S. Y., & Taeihagh, A. (2021). Adaptive governance of autonomous vehicles: Accelerating the adoption of disruptive technologies in Singapore. *Government Information Quarterly*, 38(2), 101546. https://doi.org/10.1016/j.giq.2020.101546

Thaker, J., Zhao, X., Leiserowitz, A. (2017). Media use and public perceptions of global warming in India. *Environmental Communication*, 11(3), 353-369. https://doi.org/10.1080/17524032.2016.1269824

Thomas, C. W. (1998). Maintaining and restoring public trust in government agencies and their employees. *Administration & Society*, 30(2), 166-193. https://doi.org/10.1177/0095399798302003

Tolbert, C. J., & Mossberger, K. (2006). The effects of e-government on trust and confidence in government. *Public Administration Review*, 66(3), 354-369. https://doi.org/10.1111/j.1540-6210.2006.00594.x

Trent, M., Seale, H., Chughtai, A. A., Salmon, D., MacIntyre, C. R. (2022). Trust in government, intention to vaccinate and COVID-19 vaccine hesitancy: A comparative survey of five large cities in the United States, United Kingdom, and Australia. *Vaccine*, 40, 2498-2505. https://doi.org/10.1016/j.vaccine.2021.06.048

van Noordt, C., & Misuraca, G. (2022). Artificial intelligence for the public sector: Results of landscaping the use of AI in government across the European Union. *Government Information Quarterly*, 101714. https://doi.org/10.1016/j.giq.2022.101714

Vogl, T. M., Seidelin, C., Ganesh, B., & Bright, J. (2020). Smart technology and the emergence of algorithmic bureaucracy: Artificial intelligence in UK local authorities. *Public Administration Review*, 80(6), 946-961. https://doi.org/10.1111/puar.13286

Wang, Y.-F., Chen, Y.-C., & Chien, S.-Y. (2023). Citizens' intention to follow recommendations from a government-supported AI-enabled system. *Public Policy and Administration*, 09520767231176126. https://doi.org/10.1177/09520767231176126

Wang, Y.-F., Chen, Y.-C., Chien, S.-Y., & Wang, P.-J. (2024). Citizens' trust in AI-enabled government systems. *Information Polity*, *29*(3), 293-312. https://doi.org/10.3233/IP-230065

Warren, A., Hunt, C. T., Warren, M., Bartley, A., & Manantan, M. B. (2024). Developing an AI Capability Framework for the Trilateral Security Dialogue (TSD): US, Australia, and Japan.

Williamson, O. E. (1993). Calculativeness, trust, and economic organization. *The Journal of Law and Economics*, *36*(1, Part 2), 453-486. https://www.jstor.org/stable/725485

Wirtz, B. W., Weyerer, J. C., & Kehl, I. (2022). Governance of artificial intelligence: A risk and guideline-based integrative framework. *Government Information Quarterly*, 101685. https://doi.org/10.1016/j.giq.2022.101685

Wirtz, B. W., Weyerer, J. C., & Sturm, B. J. (2020). The dark sides of artificial intelligence: An integrated AI governance framework for public administration. *International Journal of Public Administration*, 43(9), 818-829. https://doi.org/10.1080/01900692.2020.1749851

Young, M., Bullock, J., & Lecy, J. (2019). Artificial discretion as a tool of governance: a framework for understanding the impact of artificial intelligence on public administration. *Perspectives on Public Management and Governance*, 2(4), 301-313. https://doi.org/10.1093/ppmgov/gvz014

Young, M., Himmelreich, J., Bullock, J., & Kim, K.-C. (2021). Artificial intelligence and administrative evil. *Perspectives on Public Management and Governance*, 4(3), 244-258. https://doi.org/10.1093/ppmgov/gvab006

Yudkowsky, E. (2023). Pausing AI development isn't enough. We need to shut it all down. *Time*. https://time.com/6266923/ai-eliezer-yudkowsky-open-letter-not-enough/

Yuan, Q., & Chen, T. (2025). Holding AI-based systems accountable in the public sector: a systematic review. *Public Performance & Management Review*, 1-34. https://doi.org/10.1080/15309576.2025.2469784

Zuiderwijk, A., Chen, Y.-C., & Salem, F. (2021). Implications of the use of artificial intelligence in public governance: A systematic literature review and a research agenda. *Government Information Quarterly*, 38(3), 101577. https://doi.org/10.1016/j.giq.2021.101577

Zellner, A. (1962). An efficient method of estimating seemingly unrelated regressions and tests for aggregation bias. *Journal of the American Statistical Association*, 57(298), 348-368. https://doi.org/10.1080/01621459.1962.10480664